\documentclass[amsmath,amssymb]{PoS}

\usepackage{amsmath}
\usepackage{graphicx}
\usepackage{xspace}
\usepackage{color}
\usepackage{cite}
\usepackage{url}
\usepackage[utf8]{inputenc}  
\usepackage{epstopdf}

\newcommand{\bea}{\begin{eqnarray}}
\newcommand{\eea}{\end{eqnarray}}

\title{Explaining the Flavor Anomalies with a Vector Leptoquark (Moriond 2019 update)}

\author{Andreas Crivellin\thanks{I thank the organizers, especially Nazila Mahmoudi, for the invitation to Moriond QCD and the opportunity to present these results. This work is supported by a Professorship Grant (PP00P2\_176884) of the Swiss National Science Foundation.} \\
Paul Scherrer Institut, CH--5232 Villigen PSI, Switzerland \\
Physik-Institut, Universit\"at Z\"urich, Winterthurerstrasse 190, CH-8057 Z\"urich, Switzerland\\
E-mail: \email{andreas.crivellin@cern.ch}}

\author{Francesco Saturnino\thanks{I thank the organizers of the DIS2019 in Turin for giving me the opportunity to present my work, which is supported by the Swiss National Foundation under grant 200020\_175449/1. We are very grateful to Joaquim Matias and Bernat Capdevilla for providing us with the fit necessary for the $b\to s\ell^+\ell^-$ region in Fig. 2.} \\
Albert Einstein Center for Fundamental Physics, Institute
for Theoretical Physics,\\ University of Bern, CH-3012 Bern, Switzerland \\
E-mail: \email{saturnino@itp.unibe.ch}}

\abstract{Several experiments revealed intriguing hints for lepton flavor universality (LFU) violating new physics (NP) in semi-leptonic $B$ meson decays, mainly in $b \to c \tau \nu$ and $b \to s \ell^+ \ell^-$ transitions at the $3-5\,\sigma$ level. Leptoquarks (LQ) are prime candidates to address these anomalies as they contribute to semi-leptonic decays already at tree level while effects in other flavor observables, agreeing with the standard model (SM), are loop suppressed.\\
In these proceedings we review the vector leptoquark $SU(2)_L$ singlet, contained in the famous Pati-Salam model, which is able to address both $b \to c\tau \nu$ and $b \to s \mu^+ \mu^-$ data simultaneously. Due to the large couplings to tau leptons needed to account for the $b \to c \tau \nu$ data, sizable loop effects arise which we include in our phenomenological analysis. Updating our result of Ref.~\cite{Crivellin:2018yvo} with the recent measurements of LHCb~\cite{Aaij:2019wad} and BELLE~\cite{Abdesselam:2019wac,Abdesselam:2019dgh} we find an even better fit to data than before. \\[10pt] \vfill \hfill \texttt{PSI-PR-19-11}}

\FullConference{XXVII International Workshop on Deep-Inelastic Scattering and Related Subjects - DIS2019\\
		8-12 April, 2019, Torino, Italy \\
		and\\
		Moriond QCD \& High Energy Interactions, 23-30 March 2019, La Thuile, Italy}
\ShortTitle{The Pati Salam Leptoquark}

\begin{document}

\section{Introduction}
While so far the LHC has not detected any particles beyond the ones present in the Standard Model (SM), intriguing hints for LFU violation in semi-leptonic $B$-meson decays were accumulated in several (classes of) observables:

\subsection*{$b \to s \ell^+ \ell^-$}

In these flavor changing neutral current transitions, measurements of the ratios
\begin{align*}
R(K^{(*)})=\frac{\text{Br}\left[ B \to K \mu^+ \mu^- \right]}{\text{Br} \left[B \to K e^+ e^- \right]}
\end{align*}
show sizable deviations form their respective SM prediction. While the newest measurement of $R(K)$ by the LHCb collaboration~\cite{Aaij:2019wad} shows a deviation of $2.5 \, \sigma$ from the SM, the Belle result for $R(K^{(*)})$ is consistent with the SM~\cite{Abdesselam:2019wac}. However, due to the larger errors, this result also agrees with previous LHCb measurement of $R(K^{(*)})$ which deviate from the SM~\cite{Aaij:2017vbb} in the same direction as $R(K)$. Taking into account all other $b \to s \mu^+ \mu^-$ observables (like the lepton flavor universal observable $P_5'$~\cite{Aaij:2015oid}), the global fit prefers various NP scenarios above the $5 \, \sigma$ level~\cite{Capdevila:2017bsm} compared to the SM, also when the newest measurements are taken into account~\cite{Alguero:2019ptt,Aebischer:2019mlg,Ciuchini:2019usw,Arbey:2019duh}. 

In order to resolve the discrepancy in the neutral current transitions, an effect of $\mathcal{O}(10 \%)$ is required at the amplitude level. Since this flavor changing neutral current (FCNC) is suppressed in the SM as it is only induced at one loop level, a small NP contribution is already sufficient. In a global fit one finds a preference for scenarios like $C_9^{\mu \mu}=-C_{10}^{\mu \mu}$ (i.e. a left-handed current coupling to muons only)~\cite{Alguero:2019ptt}. Such an effect is naturally obtained at tree-level with the vector LQ $SU(2)$ singlet~\cite{Alonso:2015sja,Calibbi:2015kma,Fajfer:2015ycq,Barbieri:2015yvd,Barbieri:2016las,Hiller:2016kry,Bhattacharya:2016mcc,Buttazzo:2017ixm,Kumar:2018kmr,Assad:2017iib,DiLuzio:2017vat,Calibbi:2017qbu,Bordone:2017bld,Barbieri:2017tuq,Blanke:2018sro,Greljo:2018tuh,Bordone:2018nbg,Matsuzaki:2018jui,Crivellin:2018yvo,DiLuzio:2018zxy,Biswas:2018snp,Angelescu:2018tyl}. However, a $C_9^{\mu \mu}=-C_{10}^{\mu \mu}$ effect complemented by a flavor universal effect in $C_9$ gives an even better fit to data~\cite{Alguero:2018nvb,Alguero:2019ptt}. As we will see, this is exactly the pattern that arises in our model.

\subsection*{$b \to c \tau \nu$}

There are also indications for LFU violation in charged current transitions, namely in the ratios
\begin{align*}
R(D^{(*)})=\frac{\text{Br} \left[B \to D^{(*)} \tau \nu\right]}{\text{Br} \left[B \to D^{(*)} \ell \nu\right]}
\end{align*}
where $\ell = \{e,\mu\}$. While the newest measurements from Belle~\cite{Abdesselam:2019dgh} agree with the SM prediction, including previous measurements by BaBar, Belle and LHCb still yield a deviation of $3.1 \, \sigma$~\cite{Amhis:2016xyh} from the SM prediction. Furthermore there is also a measurement of the ratio $R(J/\Psi)=\text{Br}\frac{\left[B_c \to J/\Psi \tau \nu\right]}{\text{Br} \left[ B_c \to J/\Psi \mu \nu \right]}$ exceeding its SM prediction~\cite{Aaij:2017tyk}.

Also here a NP effect of $\mathcal{O}(10\%)$ is needed at the amplitude level. However, since $b \to c \tau \nu$ transitions are mediated at tree level by the exchange of a $W$ boson in the SM, the NP effect needs to be large. This means that NP should contribute at tree level with sizable couplings and at a not too high NP scale. Here, the best single particle solution is the vector LQ $SU(2)$ singlet~\cite{Alonso:2015sja,Calibbi:2015kma,Fajfer:2015ycq,Barbieri:2015yvd,Barbieri:2016las,Hiller:2016kry,Bhattacharya:2016mcc,Buttazzo:2017ixm,Kumar:2018kmr,Assad:2017iib,DiLuzio:2017vat,Calibbi:2017qbu,Bordone:2017bld,Barbieri:2017tuq,Blanke:2018sro,Greljo:2018tuh,Bordone:2018nbg,Matsuzaki:2018jui,Crivellin:2018yvo,DiLuzio:2018zxy,Biswas:2018snp,Angelescu:2018tyl} since it does not give a tree-level effect in $b\to s\nu\nu$ processes and provides a common rescaling of $R(D)$ and $R(D^*)$ with respect to the SM prediction.

\section{The Pati Salam vector leptoquark as combined solution to the anomalies}

The vector Leptoquark $SU(2)_L$ singlet with hypercharge $-4/3$, arising in the famous Pati-Salam model~\cite{Pati:1974yy}, is a prime candidate to explain both the anomalies in charged current and neutral current $B$ decays simultaneously~\cite{Alonso:2015sja,Calibbi:2015kma,Fajfer:2015ycq,Hiller:2016kry,Bhattacharya:2016mcc,Buttazzo:2017ixm,Kumar:2018kmr}. It gives a $C_9=-C_{10}$ effect in $b \to s \ell^+ \ell^-$ at tree level and at the same time a sizable effect in $b \to c \tau \nu$ without violating bounds from $b \to s \nu \nu$ and/or direct searches and does not lead to proton decay. Note that this LQ by itself is not UV complete, however several UV complete models for this LQ have been proposed~\cite{Barbieri:2015yvd,Barbieri:2016las,Assad:2017iib,DiLuzio:2017vat,Calibbi:2017qbu,Bordone:2017bld,Barbieri:2017tuq,Blanke:2018sro,Greljo:2018tuh,Bordone:2018nbg,Matsuzaki:2018jui,Cornella:2019hct}.

For the purpose of our phenomenological analysis, let us consider a model where we simply extend the SM by this LQ. Its interaction with the SM particles is given by the Lagrangian
\begin{align*}
\mathcal{L}_{V_1}=\kappa_{fi}^L \overline{Q_f} \gamma_\mu L_i V_{\mu}^{1^\dagger} + h.c.\ , 
\end{align*}
where $Q(L)$ is the quark (lepton) $SU(2)_L$ doublet, $\kappa_{fi}^L$ represents the couplings of the LQ to the left handed quarks (leptons) and $f$ and $i$ are flavor indices. Note that in principle couplings to right-handed SM particles are also allowed, they are however not relevant for this discussion. After electro-weak symmetry breaking, we work in the down basis, meaning that no CKM matrix elements appear in FCNC processes.

\begin{figure}[t]
\centering
\includegraphics[height=5cm]{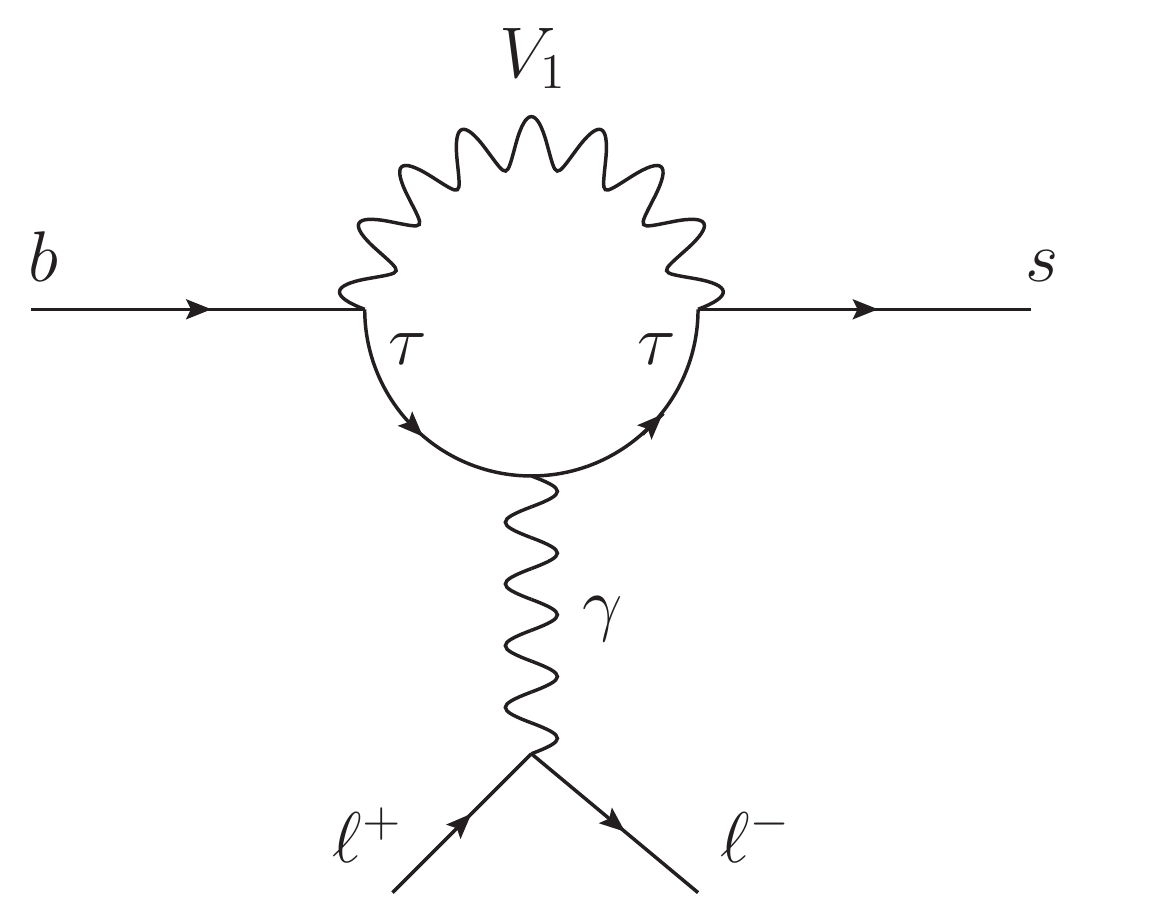}
\includegraphics[height=4cm]{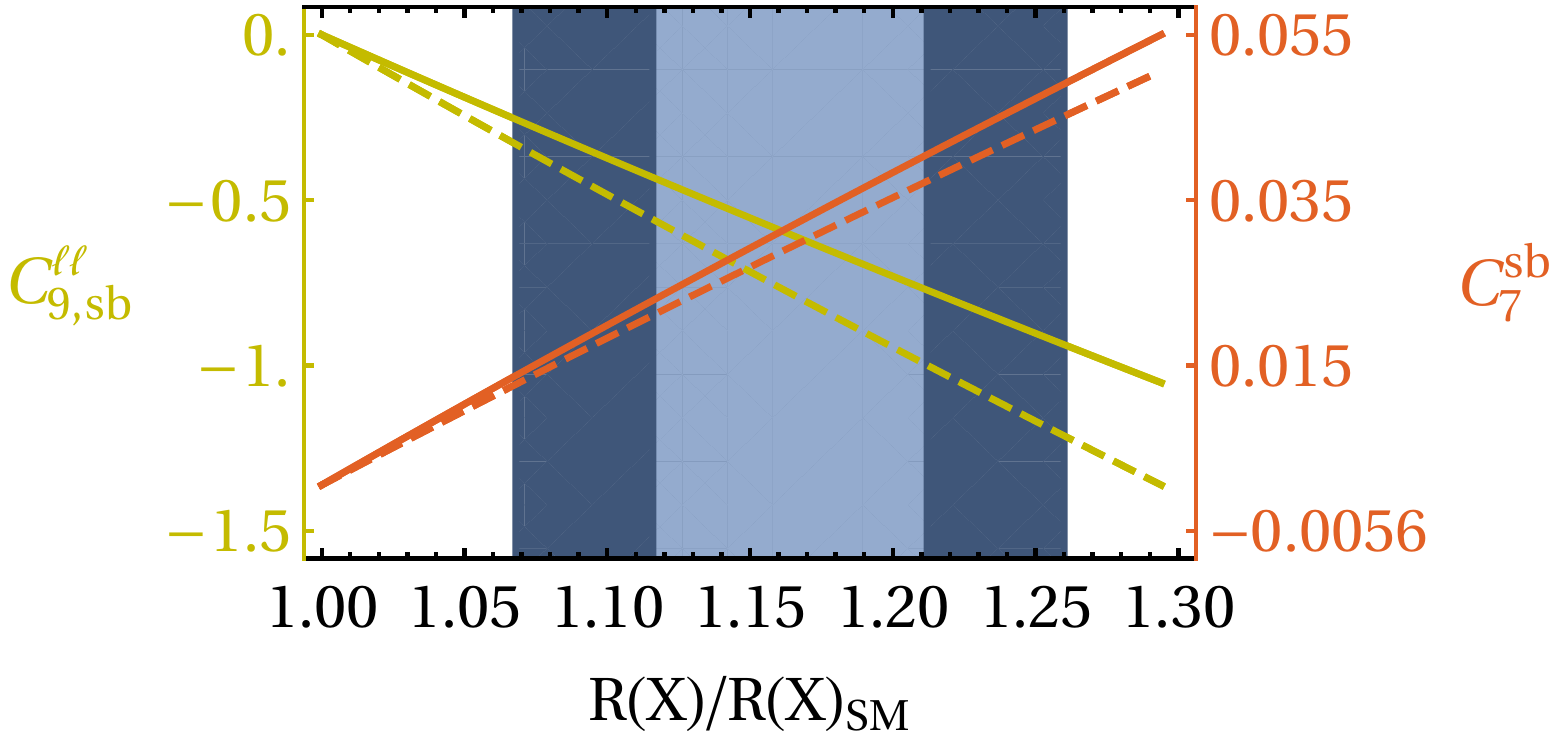}
\caption{Left: Feynman diagram depicting the loop effects induced by the $bc\tau\nu$ operator from $SU(2)$ invariance. Right: $C_{9,sb}^{\ell \ell}$ and $C_7^{sb}(\mu_b)$, generated by these loop effects, as functions of $R(D^{(*)})/{R(D^{(*)})}_{\rm{SM}}$. The solid (dashed) lines correspond to $M=1$ TeV (5 TeV)  while the (dark) blue region is preferred by $b \to c \tau \nu$ data at the $1 \, \sigma$ ($2 \, \sigma$) level, taking into account the most recent measurements. From the global fit, taking into account only lepton flavor conserving observables, we have $-1.29<C_{9,sb}^{\ell\ell}<-0.87$~\cite{Descotes-Genon:2015uva} and $-0.01<C_7^{sb}(\mu_b)<0.05$~\cite{Capdevila:2017bsm} at the $1\,\sigma$ level. Assuming an explanation of $b \to c \tau \nu$, our model predicts the right size and sign of the effect in $C_{9,sb}^{\ell \ell}$ and $C_{7}^{sb}(\mu_b)$ needed to explain $b \to s \ell^+ \ell^-$ data. }
\label{fig:lfu}
\end{figure}

We start by taking $\kappa_{23}^L$ and $\kappa_{33}^L$ as the only non-zero couplings, as they are necessary to explain $b \to c \tau \nu$ data. Here, strong effects in $b \to s \tau^+ \tau^-$ transitions~\cite{Capdevila:2017iqn} are generated which at the 1-loop level  affect $b \to s \ell^+ \ell^-$ via the Wilson coefficients $C_{9,sb}^{\ell \ell}$ and $C_7^{sb}$, as is depicted to the left in Fig.~\ref{fig:lfu}. Due to the correlation with $b \to c \tau \nu$, these Wilson coefficients can be expressed as functions of $R(D^{(*)})/{R(D^{(*)})}_{\rm{SM}}$. The Wilson coefficients' dependency on these ratios is shown in the right plot of Fig.~\ref{fig:lfu}, where the RGE evolution of $C_7^{sb}$ from the NP scale down to the $b$ quark scale is also taken into account (see Ref.~\cite{Crivellin:2019qnh}). Interestingly, assuming an explanation of $b \to c \tau \nu$ data, the effects generated in $C_{9,sb}^{\ell \ell}$ and $C_7^{sb}$ agree with the $1\,\sigma$ ranges of the model independent fit to $b \to s \mu^+ \mu^-$ data excluding LFU violating observables~\cite{Altmannshofer:2014rta,Descotes-Genon:2015uva}.

\begin{figure}[t]
\centering
\includegraphics[height=7cm]{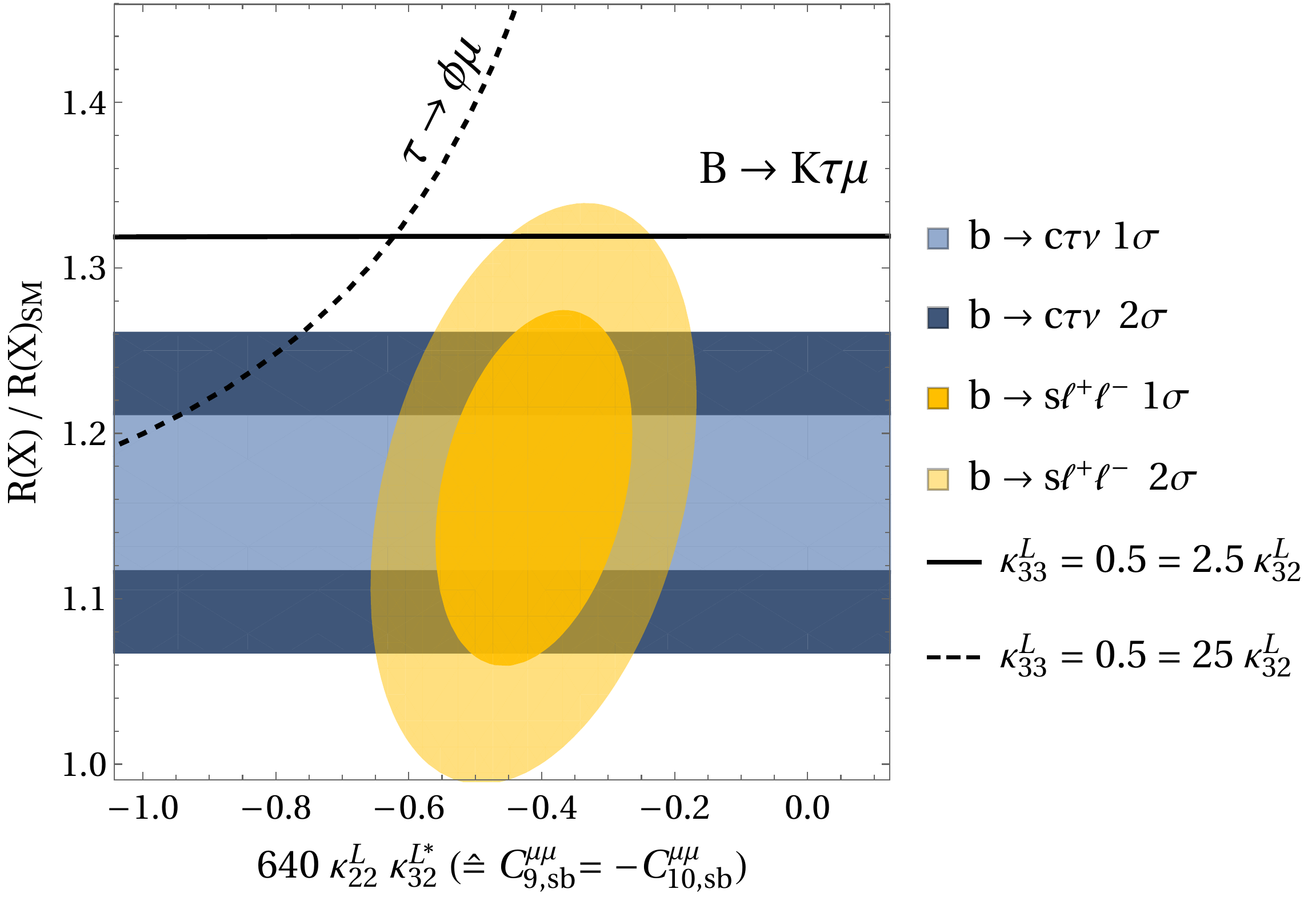}
\caption{Allowed (colored) regions in the $C_{9,sb}^{\mu \mu}=-C_{10,sb}^{\mu \mu}$ ($\equiv 640 \kappa_{22}^L \kappa_{32}^{L*}$) -- $R(X)/R(X)_{\rm{SM}}$ plane for $M=1$ TeV and $X=\{D,D^*\}$ at the $1\,\sigma$ and $2 \, \sigma$ level. The region above the black dashed (solid) line is excluded by $\tau \to \phi \mu$ ($B \to K \tau \mu)$) for $\kappa_{33}^L=0.5=25 \kappa_{32}^L$ ($\kappa_{33}^L=0.5=2.5 \kappa_{32}^L$). The bound from $\tau \to \phi \mu$ ($B \to K \tau \mu$) depends on $\kappa_{33}^L$ and $\kappa_{32}^L$ and gets stronger if $\kappa_{32}^L$ gets smaller (larger). That is, for $\kappa_{33}^L=0.5$ and $2.7 \lessapprox \kappa_{33}^L/\kappa_{32}^L \lessapprox 27$, the whole $2 \, \sigma$ region preferred by $b \to c \tau \nu$ and $b \to s \ell^+ \ell^-$ data is consistent with these bounds. Note that we used the most recent experimental results for both the $b \to c \tau \nu$ and $b \to s \ell^+ \ell^-$ transitions, therefore updating our analysis in Ref.~\cite{Crivellin:2019qnh}.}
\label{fig:b_sll}
\end{figure}

Now we also allow $\kappa_{32}^L$ and $\kappa_{22}^L$ to be non-zero, generating a tree level effect in $b \to s \mu^+ \mu^-$ which is necessary to account for the LFU violating observables as well. In Fig.~\ref{fig:b_sll} we show the allowed (colored) regions from $b \to s \mu^+ \mu^-$ and $b \to c \tau \nu$ as well as the exclusions from $b \to s \tau \mu$ and $\tau \to \phi \mu$. A simultaneous explanation of the anomalies is perfectly possible since the colored regions overlap and do not extend to the parameter space excluded by $b \to s \tau \mu$ and $\tau \to \phi \mu$. Interestingly, we predict a lepton flavor universal effect in $C_{9,sb}^{\ell \ell}$ and $C_7^{sb}$ in addition to a LFU violating tree-level effect of the form $C_{9,sb}^{\mu \mu} = - C_{10,sb}^{\mu \mu}$ in muonic channels only. This means that the effect of NP compared to the SM is expected to be larger in lepton flavor universal observables like $P5'$ relative to LFU violation observables as $R(K^{(*)})$, which is in perfect agreement with global fit scenarios~\cite{Alguero:2019ptt}. In fact, the agreement is even better after the inclusion of the new measurements of BELLE and LHCb.

\end{document}